\documentclass[preprint,5p,times,twocolumn,sort&compress]{elsarticle}

\usepackage{amssymb}

\usepackage{amsmath}

\usepackage{color}

\def\be{\begin{equation}}
\def\ee{\end{equation}}
\def\bea{\begin{eqnarray}}
\def\eea{\end{eqnarray}}
\newcommand{\lra}{\leftrightarrow}
\newcommand{\eps}{\epsilon}
\newcommand{\pa}{\partial}
\newcommand{\reef}[1]{(\ref{#1})}
\newcommand{\la}{\langle}
\newcommand{\ra}{\rangle}
\makeatletter
\def\ps@pprintTitle{%
     \let\@oddhead\@empty
     \let\@evenhead\@empty
     \let\@oddfoot\@empty
     \let\@evenfoot\@empty
     }
\makeatother

\begin{document}

\begin{frontmatter}


\title{On the Supersymmetrization of Galileon Theories in Four Dimensions}

\author{Henriette Elvang\corref{corauth}}
\ead{elvang@umich.edu}
\author{Callum R.~T.~Jones\corref{primcorauth}}
\ead{jonescal@umich.edu}
\author{Marios Hadjiantonis\corref{corauth}}
\ead{mhadjian@umich.edu}
\author{Shruti Paranjape\corref{corauth}}
\ead{shrpar@umich.edu}

\cortext[primcorauth]{Primary corresponding author}
\cortext[corauth]{Corresponding author}

\address{
Leinweber Center for Theoretical Physics,\\
Randall Laboratory of Physics, Department of Physics,\\
University of Michigan, Ann Arbor, MI 48109, USA\\
LCTP-1714
}

\begin{abstract}

We use on-shell amplitude techniques to study the possible $\mathcal{N}=1$ supersymmetrizations of Galileon theories in $3\!+\!1$ dimensions, both in the limit of decoupling from DBI and without. Our results are that (1) the quartic Galileon has a supersymmetrization compatible with Galileon shift symmetry  ($\phi \to \phi +c + b_\mu x^\mu$) 
for the scalar sector and an ordinary shift symmetry  ($\psi \to \psi + \xi$)
for the fermion sector, and it is unique at least at 6th order in fields, but possibly not beyond; 
(2) the enhanced ``special Galileon'' symmetry is incompatible with supersymmetry;
(3) there exists a quintic Galileon with a complex scalar preserving Galileon shift symmetry; 
(4) one cannot supersymmetrize the cubic and quintic Galileon while preserving the Galileon shift symmetry for the complex scalar; 
and
(5) for the quartic and quintic Galileon, we present evidence for a supersymmetrization in which the real Galileon scalar is partnered with an R-axion to form a complex scalar which only has an ordinary shift symmetry.

\end{abstract}


\end{frontmatter}


\section{Introduction and Results}
\label{sec:Introduction}
Galileon theories are scalar effective field theories (EFTs) with higher derivative self-interactions of the form 
\be 
   \label{GalInt}
   \mathcal{L} = \tfrac{1}{2}(\pa\phi)^2 + \sum_{n=3}^{D+1} g_n (\pa\phi)^2 (\pa\pa\phi)^{n-2}\,,
\ee 
where $D$ is the spacetime dimension. The couplings $g_n$ are generally independent. The characteristic feature of these models is that despite the higher derivatives, the equations of motion are only second order. As a consequence, the Galileons have a well-defined classical field theory limit, free from Ostrogradski ghosts. This feature is strongly atypical among EFTs and make Galileons attractive for model building in cosmology and beyond. The cubic Galileon originally arose in the Dvali-Gabadadze-Porrati (DGP) model \cite{Dvali:2000hr}, but Galileons appear 
in other contexts too, for example in modifications of gravity \cite{Nicolis:2008in,deRham:2010ik,deRham:2010kj}. Perhaps most significantly, Galileons emerge as subleading terms on effective actions on branes \cite{deRham:2010eu}. Here we focus on flat branes in Minkowski space, although other embeddings 
are also of interest \cite{deRham:2010eu,Goon:2010xh,Goon:2011qf}. 

A flat 3-brane placed in a $4\!+\!1$-dimensional Minkowski bulk will induce a spontaneous breaking of spacetime symmetry: $\text{ISO}(4,1)\rightarrow \text{ISO}(3,1)$. A massless Goldstone mode $\phi$ must appear in the spectrum of the $3\!+\!1$-dimensional world-volume EFT which is physically identified with fluctuations of the brane into the extra dimension. The full $\text{ISO}(4,1)$ symmetry remains a symmetry of the action, and so at leading and next-to-leading order in the derivative expansion the effective action takes the form \cite{deRham:2010eu}
\begin{equation} \label{DBIGal}
  S = \int \text{d}^4x \sqrt{-G} \left[\Lambda^4_2 + \Lambda^3_3 K[G]+\Lambda^2_4 R[G]+\Lambda_5 \mathcal{K}_{\text{GHY}}[G]\right]\,,
\end{equation}
where $G$ is the pullback of the bulk metric onto the 3-brane world-volume. The leading term with coupling $\Lambda_2^4$ (the brane tension) gives the Dirac-Born-Infeld (DBI) action, while the remaining terms are built from the extrinsic curvature, intrinsic curvature and Gibbons-Hawking-York (GHY) boundary terms.  The $\Lambda_i$ are in general arbitrary mass scales. The resulting action is the {\em DBI-Galileon} \cite{deRham:2010eu}: its leading part is DBI and the subleading terms are cubic, quartic, and quintic in $\phi$.\footnote{The boundary terms $K[G]$ and $\mathcal{K}_{\text{GHY}}[G]$ are only available when the brane is considered an end-of-the-world brane and they are responsible for the odd-powered 
$\phi$-interactions.} 
The non-DBI interaction terms in (\ref{DBIGal}) are the $4\!+\!1$-dimensional Lovelock invariants that give the characteristic second-order equations of motion.

 The {\em Galileon models} (\ref{GalInt}) correspond to a decoupling limit in which the 3-brane tension $\Lambda_2^4 \rightarrow \infty$, but the ratios
\begin{equation}
 \label{decouple}
  g_3 = \frac{\Lambda^3_3}{\Lambda^6_2}, \;\;\;g_4 = \frac{\Lambda^2_4}{\Lambda^8_2}, \;\;\;g_5 = \frac{\Lambda_5}{\Lambda^{10}_2}\,, 
\end{equation}
are held fixed. The only part of DBI that survives is the canonical kinetic term for $\phi$.\footnote{When decoupled from DBI, Galileons violate the null-energy condition and for that reason they have received attention as models for cosmological bounces. However, without the leading DBI terms, the Galileon theories cannot arise as the low-energy limit of a UV complete theory \cite{Adams:2006sv}.}

The Galileons and the DBI-Galileons both enjoy a non-trivial extended shift symmetry of the form 
\be
\label{galsym}
\phi \to \phi + c+ b_\mu x^\mu + \ldots\,,
\ee
where $c$ is a constant, $b_\mu$ is a constant vector, and $x^\mu$ is the spacetime coordinate. The ellipses stand for possible field-dependent terms, which will not play a role for us here. These symmetries arise from the spontaneously broken symmetry 
 generators \cite{Goon:2010xh}: the constant shift from the broken bulk translation and the $x^\mu$-shift from the broken Lorentz rotation.

The quartic Galileon ($g_3=g_5=0$) is sometimes called the {\em special Galileon} \cite{Cheung:2014dqa,Hinterbichler:2015pqa} because it has a further enhanced shift symmetry 
\be
\label{specgalsym}
\phi \to \phi + s_{\mu\nu} x^\mu x^\nu + \ldots \,,
\ee
where the constant tensor $s_{\mu\nu}$ is symmetric and traceless \cite{Hinterbichler:2015pqa}. 
This is an accidental symmetry that  occurs  only in the decoupling limit from DBI. 

In this paper, we address the question of supersymmetrization of Galileon theories in $3\!+\!1$ dimensions, both in the context of DBI-Galileons and the decoupled Galileons.  Based on the brane construction, one expects that the quartic DBI-Galileon can be supersymmetrized, in particular, there should exist an $\mathcal{N}=4$ supersymmetrization corresponding to the effective action for a D3-brane in 9+1-dimensional Minkowski space. It is less obvious that supersymmetry would survive the decoupling limit or if the cubic or quintic  (DBI-)Galileons 
can be supersymmetrized. An explicit $\mathcal{N}=1$ superfield construction of the quartic Galileon was  presented in \cite{Farakos:2013zya}. 
We will construct an $\mathcal{N}=1$ quartic Galileon and comment on its uniqueness. 
In the literature, a supersymmetrization of the cubic Galileon was proposed, but it suffered from ghosts \cite{Koehn:2013hk}.  By a field redefinition, any cubic Galileon is equivalent to the quartic and quintic Galileon with related couplings, so we address supersymmetrization of the cubic Galileon via the quintic. Before describing our approach, let us briefly comment on the super-algebra.

\subsection{Symmetry algebra}

The Poincare algebra can be extended \cite{Goon:2012dy,Hinterbichler:2015pqa} with the translation generator  $C$  ($\delta_C \phi = 1$),  the Galileon shift generator $B_\mu$ ($\delta_B \phi = x^\mu$), and  the symmetric traceless generator $S_{\!\mu\nu}$ of the special Galileon transformations \reef{specgalsym}. 

Being agnostic about the origin of a Galileon extension of the super-Poincare algebra, at the minimum we might demand the closure of the extended super-translation sub-algebra with generators $P_\mu$, $Q$, $\bar{Q}$, $C$, and $B_\mu$ (plus $S_{\!\mu\nu}$ for the special Galileon), as well as a second set of fermionic generators $S$ and $\bar{S}$ associated with spontaneously broken supersymmetry.  The latter are required by the algebra. 
Among the new commutator relations, we must have (schematically)
\be
  \begin{split}
 &
 [P_\mu, B_\nu] \sim \eta_{\mu\nu} C\,,~~~
 [P_\rho, S_{\!\mu\nu}] \sim \eta_{\mu\rho} B_\nu + \eta_{\nu\rho} B_\mu \,,\\
 &
 [B_\mu, Q] \sim  \sigma_\mu(\bar{Q} + \bar{S})\,,~~~
 [S_{\!\mu\nu}, Q] =  0\,.~~~
 \end{split}
\ee
The last vanishing commutator follows from the fact that  $[S_{\!\mu\nu}, Q]$ must be a linear-combination of fermionic generators, but there are no tensor structures available that can make it symmetric and traceless. 
Now, consider the Jacobi identity
\be
  \label{SJacobi}
  [S_{\!\mu\nu},\{ Q, \bar{Q}\}] = \{[S_{\!\mu\nu}, Q] ,\bar{Q}\}+ \{Q, [S_{\!\mu\nu}, \bar{Q}]\}\,.
\ee
The RHS vanishes, but using $\{ Q, \bar{Q}\} \sim P$ the LHS gives a non-vanishing linear combination of $B_\mu$-generators. Therefore  
the algebra does not close consistently. This indicates 
that there is no supersymmetrization of the special Galileon that also preserves the enhanced symmetry \reef{specgalsym}.  
Replacing $S_{\!\mu\nu}$ by $B_\mu$ in the Jacobi identity \reef{SJacobi} gives $C$ on the LHS. The RHS can match this if $\{ Q, S\} \sim C$. There does not appear to be any inconsistency extending the super-translation algebra with the Galileon generators $C$ and $B_\mu$. 
Indeed, such an algebra follows from the scenario of bulk supersymmetry spontaneously broken to $\mathcal{N}=1$ on the 3-brane.\footnote{The decoupling limit \reef{decouple} induces an \.{I}n\"{o}n\"{u}-Wigner contraction of the original $\text{ISO}(4,1)$ symmetry algebra in the direction transverse to the 3-brane. The resulting algebra $\mathfrak{Gal}(4,1)$ is a cousin of the familiar Galilean algebra of non-relativistic mechanics. In the decoupling limit \reef{decouple}, the extended shift symmetry (\ref{galsym}) arises from the non-linear realization of the coset $\mathfrak{Gal}(4,1)/\text{ISO}(3,1)$. The recent work \cite{Roest:2017uga} extends this construction to include the supercharges. An earlier version of the algebra is in \cite{Bagger:1994vj}.
}

These algebraic arguments constrain the form of the symmetry as realized on the classical fields and are suggestive but formally problematic when extended to the quantum theory. In general, spontaneously broken symmetries do not possess well-defined Noether charges as operators on a Hilbert space.\footnote{The algebra constructed in \cite{Roest:2017uga} is of the former kind. In this case even the classical Poisson algebra will differ from the algebra realized on the fields by the appearance of central terms \cite{Toppan:2001qb}.   } As demonstrated in \cite{PhysRevLett.16.408.2}, the infinite volume improper integral of the Noether charge density operators of spontaneously broken symmetries do not converge in the weak operator topology.  Furthermore, the second $S$-type supersymmetry is necessarily spontaneously broken and satisfies a current algebra with tensor central charges which cannot be integrated to a consistent charge algebra in infinite volume. (See \cite{Dumitrescu:2011iu} for a related discussion.) 
It is difficult to draw convincing conclusions from an algebra which formally does not exist.

Nonetheless we will find that the properties suggested by the algebraic arguments do indeed hold as properties of the scattering amplitudes and can be argued for in a mathematically satisfactory way. 
In the following, we outline the strategy of using on-shell amplitudes to assess the existence of effective field theories with linearly- and non-linearly realized supersymmetry. 
We then apply these methods to prove each of claims in the Abstract.

\subsection{Approach}

The precise form of the extended shift symmetry (\ref{galsym}) is parametrization-dependent. The consequences for physical observables, however, are not. From such non-linear symmetries follow universal soft theorems which must be satisfied by all on-shell scattering amplitudes. Similarly the form of a supersymmetry transformation on the fields appearing in the effective action depends on the parametrization, but the on-shell supersymmetry Ward identities relating the scattering amplitudes do not. Both of these properties will be used to place constraints on any possible effective field theory with both Galileon extended shift symmetry and supersymmetry. 

The traditional approach to the supersymmetrization of a bosonic theory is to work with the Lagrangian and either supersymmetrize the terms directly or to employ superfields. If one is not successful, it can be hard to strictly rule out the existence of supersymmetrizations. And if one is successful, can one know with certainty if the answer is unique?
Finally, higher-derivative theories are haunted by questions of field redefinitions that can confuse matters of whether two Lagrangians are physically equivalent. A more systematic approach uses the methods of non-linear realizations of Callan, Coleman, Wess and Zumino \cite{Coleman:1969sm,Callan:1969sn} and Volkov \cite{Volkov:1973vd}, but these are based on specific symmetry-breaking patterns. We avoid all the above issues by addressing the question of supersymmetrizations from the point of view of on-shell amplitudes. This will allow us to give precise exclusion statements and give evidence for existence of supersymmetric Galileons.

Technical details are discussed with much further rigor in the forthcoming paper \cite{ourlongpaper}, which will also include analytic results for scattering amplitudes of Galileons and their superpartners. 

The evidence for new (supersymmetric) theories presented in this paper is based on on-shell amplitudes. If one engineers the associated Lagrangians, it is not clear if their  equations of motion will remain second-order and hence free of Ostrogradski ghosts. This would be interesting to address in future work.

\section{Method}
\subsection{Lagrangian  $\leftrightarrow$ amplitudes}
If a local Lagrangian has an $n$-field interaction term which is not a total derivative and not proportional to the equation of motion of any of its fields, then it has an associated $n$-particle on-shell matrix element which has no poles, i.e.~it is a polynomial in the kinematic variables. The number of independent kinematic polynomials that obey all the symmetries of the underlying theory tells us how many such independent Lagrangian terms there are.
Thus, to assess if a local interaction term exists with certain symmetries, we impose these symmetry constraints on the most general ansatz for the amplitude. If no such amplitude exists, it means that no such independent local interaction term exists. If an amplitude exists, it is evidence that the theory may exist and we can further characterize the properties of the theory using explicitly computed scattering amplitudes. 

Shift symmetries, such as \reef{galsym}, of the Lagrangian manifest themselves as Adler zeroes of  the amplitudes. 
When a single external momentum is taken soft, $p_i \to \eps\,p_i$ with $\epsilon \to 0$, the tree amplitude vanishes as
\be
  A_n \sim
  \eps^\sigma ,
\ee
where $\sigma$ is an integer. In particular, the  ordinary shift symmetry $\phi \to \phi + c$ gives a soft theorem with $\sigma =1$, the Galileon symmetry $\phi \to \phi + c_\mu x^\mu$ gives $\sigma =2$, and the special Galileon shift \reef{specgalsym} gives $\sigma=3$.

\subsection{Spinor helicity}
We use spinor helicity formalism to encode particle kinematics in $3\!+\!1$ dimensions. 
The on-shell  momenta of the external massless particles can be expressed in terms of commuting angle and square spinors $|p\ra$ and $|p]$ via
\be
  p_\mu (\sigma^\mu)_{a\dot{b}} 
  =
  p_{a\dot{b}} 
  =
   - |p]_a \la p |_{\dot{b}}\,.
\ee
Spinor indices are raised and lowered with the 2-index Levi-Civita symbol, so Lorentz-invariant contractions are antisymmetric:
\be
  \la p q \ra = \la p |_{\dot{a}} | q \ra^{\dot{a}} = - \la q p \ra \,,
  ~~~~~
  [ p q ] = [ p |^a | q ]_a = - [ q p ] \,.
\ee 
We note the following useful relation between Mandelstam variables and spinor brackets
\be
  s_{ij} = -(p_i + p_j)^2 = -2p_i \cdot p_j = - \la i j \ra [ij]\,.
\ee 
When the momentum is real, angle and square spinors are related by complex conjugation. Often it is useful to analytically continue to complex momenta and the angles and squares are then independent. 

External line Feynman rules are  also 
written in terms of spinor variabes. 
The wavefunction for a positive (negative) helicity massless fermion with momentum $p$ is $|p]$ ($|p\ra$). A review of spinor helicity and expressions for polarization vectors can be found in \cite{Elvang:2013cua,Elvang:2015rqa}. 

Little group scaling $( |i\ra, |i] ) \to ( t |i\ra, t^{-1}|i] )$ leaves the on-shell momentum invariant and the amplitude scales homogenously as $A_n \to t^{-2h_i} A_n$ where $h_i$ is the helicity of particle $i$. 

To avoid  trivially  vanishing results due to little group scaling, we take the soft limit $p_i \to \epsilon p_i$ differently for non-negative and negative helicity particles:
\be
  \begin{split}
  h_i \ge 0\!:&~~~~~|i\ra  \to \eps |i\ra \,,~~~~|i]  \to  |i]\,,\\
  h_i < 0\!:&~~~~~|i\ra  \to  |i\ra \,,~~~~|i]  \to \eps |i]\,.
  \end{split}
\ee
We actually take the soft limit in a way that preserves momentum conservation, see  \cite{Elvang:2016qvq}. 

\subsection{Scalar example}
Let us apply this to an example to illustrate the ideas. Suppose we would like to examine the existence of quintic interaction terms with Galileon symmetry \reef{galsym}. Such terms would take the schematic form $g \pa^{2m} \phi^5$ and the mass-dimension of the coupling therefore has to be $[g]=-(2m+1)$. The most general ansatz for the  5-scalar amplitude $A_5(\phi\phi\phi\phi\phi)$ must be a Bose-symmetric degree $m$ polynomial in the Mandelstam variables $s_{ij}$ (i.e.~$2m$ in powers of momentum). 
It is straightforward to list the possible polynomials for small values of $m$:\footnote{Throughout this paper, we use crossing symmetry to take amplitudes to have all outgoing particles. We consider only  tree level.
}
\be
\begin{array}{lll}
  \pa^{2m} \phi^5 &A_5(\phi\phi\phi\phi\phi) \\
  m=0 &  1\\
  m=1 & 0~ \text{~by mom.cons.}\\
  m=2 & \sum_{i<j} s_{ij}^2 \\
  m=3 & \sum_{i<j} s_{ij}^3 \,.
\end{array}
\ee
None of these have vanishing soft limits, so there can be no quintic single scalar Lagrangians at these orders with shift symmetry. 

The quintic Galileon has coupling of mass-dimension $-9$, i.e.~$m=4$, so that means the terms in the Lagrangian have 8 derivatives distributed on the 5 scalar fields. There are two polynomials that are linearly independent modulo momentum conservation, so we can write 
\be
 \label{A5realGal}
  A_5(\phi\phi\phi\phi\phi) = 
  c_1 \sum_{i<j} s_{ij}^4
  + c_2 \Big( \sum_{i<j} s_{ij}^2 \Big)^2\,.
\ee
This means that there are two independent ways that 8 derivatives can be distributed on 5 identical scalar fields in the Lagrangian and they are not related by partial integration or application of the equation of motion.  For generic coefficients $c_1$ and $c_2$, this amplitude does not go to zero when a momentum is taken soft. This only happens when $c_2=-c_1/4$, in which case one actually finds that the amplitude vanishes as $O(\eps^2)$. This solution is the unique 5-point Galileon. 

Starting at 10 derivatives (coupling dimension $-11$), each of the five fields $\phi$ can be dressed with two or more derivatives and then invariance under \reef{galsym} is trivial. The quintic Galileon is therefore rather special: it is the only quintic interaction  of a single real scalar field that is non-trivially invariant under the extended  
shift symmetry \reef{galsym}. 

\subsection{Supersymmetry Ward identities}
The supercharges $Q$ act on the on-shell states of a chiral multiplet as
\be
  [Q, \psi ] = |p] Z\,,~~
  [Q, \bar\psi] = 0\,,~~
  [Q, \bar Z ] = |p] \bar\psi\,,~~
  [Q, Z ] =0\,,
\ee
where the brackets are graded and we use the fields as shorthand to denote the on-shell states. For brevity, we use $\psi$ and  $\bar \psi$ to denote a fermionic external state with positive and negative helicity, respectively.

The supersymmetry Ward identities follow from $Q$'s annihilation of the vacuum,
\be
  0= \la [Q, \Phi \dots \Phi]\ra = \sum_i \la \Phi \dots [Q,\Phi_i] \dots \ra \,,
\ee
where $\Phi$ stands for any state in the chiral multiplet.
For example $0=\la [Q, \psi Z\dots Z ]\ra = |1] \la ZZ \dots Z \ra $ tells us that the amplitude with all $Z$ external states must vanish. Likewise 
\be
  0=\la [Q,  \psi \bar Z Z\dots Z]\ra 
  = 
  |1] \la Z \bar Z Z \dots Z \ra - |2] \la \psi \bar \psi Z \dots Z \ra
\ee
implies that both these amplitudes must vanish, as can be seen from dotting in $[1|$ or $[2|$.
Thus, the only complex scalar amplitudes that can be compatible with supersymmetry have at least two $Z$'s and two $\bar Z$'s. The analogue statement for gluon amplitudes in super-Yang Mills theory is that the helicity-violating amplitudes $++\ldots+$ and $-+\ldots+$ vanish. 

\subsection{Supersymmetry example}
The supersymmetry Ward identity for 4-particle amplitudes of a massless chiral multiplet with a complex scalar $Z$ and a Weyl fermion $\psi$ takes the form 
\be
  \label{4ptsusyWI}
  A_4(\psi \bar{\psi} \psi \bar{\psi}) 
  =
  \frac{[13]}{[23]} A_4(Z \bar{Z} \psi \bar{\psi})
  = 
  \frac{[13]}{[24]} A_4(Z \bar{Z} Z \bar{Z})\,.
\ee

To illustrate the application, consider 4-particle amplitudes with couplings of dimension $-4$. Four-particle tree amplitudes  
must be dimensionless, so a local amplitude at this order must be a linear combination of the independent 2nd order Mandelstam polynomials. Taking Bose symmetry into account, there are two possible such terms for the amplitude 
$A_4(Z \bar{Z} Z \bar{Z}) = b_1 t^2 + b_2 s u$.  Only the $t^2$-term 
 is compatible with supersymmetry. 
To see this, let us consider the 4-fermion amplitude. 
Little group scaling and antisymmetry under exchanges of identical fermions tell us that $A_4(\psi \bar{\psi} \psi \bar{\psi})$ must be equal to $\la 24 \ra [13]$ times a Mandelstam polynomial of degree 1 symmetric under $1 \lra 3$ and $2 \lra 4$  (the spinor brackets account for two mass-dimensions): 
there is only one option, namely
\be
  A_4(\psi \bar{\psi} \psi \bar{\psi}) = b_1' \la 24 \ra [13] t\,.
\ee
Now using the supersymmetry Ward identities, we find
\be
  A_4 (Z \bar{Z} Z \bar{Z}) = \frac{[24]}{[13]}A_4(\psi \bar{\psi} \psi \bar{\psi}) 
   = b_1'  \frac{[24]}{[13]}  \la 24 \ra [13] t
   = - b_1' t^2\,. 
\ee
Thus we see that only the case of $b_1 = - b_1'$ and $b_2=0$ is compatible with supersymmetry. This is the unique solution that corresponds to the $\mathcal{N}=1$ supersymmetric theory of DBI coupled to Akulov-Volkov, the effective theory of Goldstinos.

\subsection{Soft subtracted recursion}
We use {\em soft subtracted recursion relations} to test the soft behavior. This method was introduced in \cite{Kampf:2013vha,Cheung:2014dqa,Cheung:2015cba,Cheung:2015ota,Cheung:2016drk,Luo:2015tat}. We outline how it works, but leave the details of our analysis for \cite{ourlongpaper}.

Consider deforming the $n$ complex 
external momenta $p_i$ as
\be
\hat{p}_i^\mu = p_i^\mu (1-a_i z)\,. 
\ee
The shift parameters $a_i$ must be chosen subject to the constraint of momentum conservation, $\sum_{i=1}^n p_i^\mu a_i =0$. As a function of these shifted momenta (which are on-shell too), we 
 write the tree-level amplitude $\hat{A}_n(z)$. Consider the integral 
\be
  \label{contourint}
  \oint dz \, \frac{\hat{A}_n(z)}{z\, F(z)} 
  ~~~\text{with}~~~
  F(z) = \prod_{i=1}^n (1 - a_i z)^{\sigma_i} \,,
\ee
where the contour surrounds all finite poles in the $z$-plane.  This integral vanishes provided there are no poles at $z=\infty$; a sufficient condition can be stated  in terms of the mass-dimensions of the couplings $[g_v]$ that appear as vertices in the $n$-point amplitude \cite{ourlongpaper}:   
\be
  \label{reccrit}
  4 - n - \sum_v [g_v] - \sum_{i=1}^n s_i - \sum_{i=1}^n \sigma_i < 0\,,
\ee
where $s_i \ge 0$ is the spin (not helicity) of particle $i$ and $\sigma_i$ are the non-negative integers that appear in the definition of the function $F(z)$ in \reef{contourint}.

If the criterion \reef{reccrit} holds, then the residue at $z=0$, which gives the unshifted amplitude, equals minus the sum of all the other finite-$z$ residues. These come in two classes: from simple poles in the amplitude and from the poles at $z=1/a_i$. The former we can calculate easily since the amplitude factorizes on its physical poles. To deal with the latter, note that taking $z\to 1/a_i$ is simply a single soft limit of the shifted amplitude  $\hat{A}_n(z)$ (assuming all the $a_i$ be distinct). Hence, if $A_n$ vanishes in the single soft limit at the rate $\sigma_i$ for particle $i$,  there are no poles in \reef{contourint} at $z = 1/a_i$. That means the amplitude is fully constructible from its factorization into lower-point amplitudes. The purpose of the function $F(z)$ is to improve the large-$z$ behavior, and this is absolutely needed in effective field theories with higher-derivative couplings. 

We illustrate the method with an example. Consider again the general ansatz $A_4(Z \bar{Z} Z \bar{Z}) = b_1 t^2 + b_2 s u$ for a complex scalar amplitude in an EFT whose quartic coupling has dimension $-4$. In the Lagrangian approach, one would calculate the 6-particle amplitude $A_6(Z \bar{Z} Z \bar{Z} Z \bar{Z})$ from pole diagrams with the 4-point interactions and possible 6-point contact terms in that theory. To calculate $A_6$ recursively, we note that it factorizes into two 4-point amplitudes $A_4(Z \bar{Z} Z \bar{Z})$, so the criterion  \reef{reccrit} is $4-6-2(-4)-0- 6\sigma= 6(1-\sigma) < 0$, i.e.~soft subtracted recursion is valid for computation of $A_6$ if $\sigma \ge 2$. 
Now \emph{assume} that $\sigma=2$ is a property of the 6-particle amplitude and calculate $A_6$ via the soft subtracted recursion relation. Note that this is the soft behavior of the complex scalar, meaning that both its real and imaginary parts  are assumed to have the symmetry \reef{galsym}. If the assumption of $\sigma=2$ holds true, the  recursive result 
for $A_6$ must be independent of the $a_i$'s.\footnote{In $3\!+\!1$ dimensions, the constraint of momentum conservation has $n-4$ solutions. For all $n$ there is a trivial solution with all $a_i$ equal, however this solution cannot be used to defined subtracted recursion relations. If the external momenta are restricted to a 3-dimensional subspace then the number of solutions increases to $n-3$. Demanding that the candidate amplitude is independent from the choice of $a_i$ in the restricted kinematics often gives stronger constraints.}
  Hence, if $a_i$-independence fails, the amplitude cannot have $\sigma=2$ and therefore there cannot be a corresponding Lagrangian invariant under Galileon symmetry. For the particular case at hand, one finds that $a_i$-independence works only when $b_2=0$. The conclusion is that there is definitively no EFT with Galileon symmetry \reef{galsym} for the  complex scalar when $b_2\ne0$. 
  On the other hand, we have positive evidence in favor of  such a theory whose 4-point interaction gives $A_4(Z \bar{Z} Z \bar{Z}) =  b_1 t^2$. This theory is already well-known: it is the complex scalar DBI theory --- and we learned previously that it is compatible with supersymmetry.

If we assume $\sigma=3$, we will find that recursion fails (the result is $a_i$-dependent). This tells us that (the complex scalar) DBI theory cannot be invariant under the special Galileon symmetry \reef{specgalsym}. 

To summarize, the soft subtracted recursion relations allow us to efficiently rule out existence of effective field theories with given fundamental interaction vertices and spontaneously broken symmetries. They also provide evidence (though not a proof) of existence of theories and explicit results for the scattering amplitudes. 

\section{Results I}

The first step towards supersymmetry is to combine the Galileon with another scalar to form a complex scalar $Z$ of a chiral supermultiplet. We will consider two cases in this paper. In this section,   both the real and imaginary part of $Z$ have the extended shift symmetry \reef{galsym}. In Sec.~\ref{s:raxion}, we relax this condition.

\subsection{Complex scalar quintic Galileon}
\label{complxQuinticGal}

 Multi-scalar Galileon theories were constructed in \cite{Hinterbichler:2010xn} from the effective action of a 3-brane in a bulk space with $n$ transverse directions. The actions in \cite{Hinterbichler:2010xn} have $n$ scalars which are Goldstone modes of each of the spontaneously broken translational symmetries and they all have extended shift symmetry \reef{galsym}. The models inherit SO$(n)$ symmetry from the bulk, in particular that means there are only even-powered interactions. This may seem to doom a quintic Galileon with more than one scalar; however, we now give evidence for the existence of a complex scalar quintic model that breaks U$(1)$=SO$(2)$, but still has symmetry \reef{galsym} for both real scalars.

To  potentially be compatible with supersymmetry, the complex scalar 5-point amplitude must be of the form $A_5(Z \bar Z Z \bar Z Z )$ (and its conjugate). The coupling has mass-dimension $-9$, and the interaction terms have 8 derivatives. Using that the 5 momenta must satisfy momentum conservation, one finds (using Mathematica to generate the polynomial basis) that there are ten independent Lorentz-invariant contractions of 8 momenta satisfying Bose symmetry under exchanges of identical states $\{1 \lra 3 \lra 5\}$ and $\{ 2 \lra 4\}$. Of these ten, nine are polynomials of degree 4 in the Mandelstam variables, whereas the tenth is parity odd and proportional to the Levi-Civita symbol.

Imposing the Galileon symmetry in the form of the required $\sigma=2$ soft behavior of a general linear combination of these ten basis polynomials selects one unique answer:
\be
  \label{5ptcomplexGal}
  \begin{aligned}
  A_5(Z \bar Z Z \bar Z Z) & = 
  c_1 \bigg( \sum_{i<j} s_{ij}^4 - \frac{1}{4} \Big( \sum_{i<j} s_{ij}^2 \Big)^2  \bigg)\\
  & = 48 c_1 \Big ( \epsilon_{\mu \nu \rho \sigma} p_1^\mu p_2^\nu p_3^\rho p_4^\sigma \Big )^2\,.
  \end{aligned}
\ee
This amplitude is equal to the real 5-point Galileon \reef{A5realGal},  however, at higher-point these models give distinct amplitudes. For example, using soft subtracted recursion relations (which are valid by \reef{reccrit}) to obtain the  8-point amplitudes involves 
$\frac 1 2 \left ( \begin{smallmatrix} 8 \\ 4 \end{smallmatrix} \right ) = 35$ factorization diagrams for the real scalar case, while for the complex scalar case there are actually two types of 8-point amplitudes, $A_8(Z \bar Z Z \bar Z Z \bar Z Z \bar Z)$ and $A_8(Z \bar Z Z \bar Z Z \bar Z Z  Z)$. The former has 52 diagrams (of two different types) and the latter has 30 diagrams. We have computed these three 8-point amplitudes and verified that they are distinct. We conclude that 
\begin{quote}
{\em this is significant evidence in favor of the 
 existence of a 5-point Galileon whose complex scalar has Galileon symmetry \reef{galsym}. }
\end{quote}
Note that this model necessarily breaks any $U(1)$ symmetry acting on the scalars.  Next we show that this quintic model is not compatible with supersymmetry.

\subsection{Quintic Galileon \& supersymmetry, part 1}
One necessary condition for supersymmetry is the Ward identity 
\be
 \label{5ptsusyWI}
 A_5(Z \bar Z Z \bar \psi \psi) = -\frac{[25]}{[24]}A_5(Z \bar Z Z \bar Z Z)\,.
\ee 
The amplitude on the LHS must come from a local interaction term in the Lagrangian that arises from the supersymmetrization of the five-scalar term. So $A_5(Z \bar Z Z \bar \psi \psi)$ must be local, i.e.~it cannot have any poles.\footnote{One might worry about contributions from pole diagrams involving Yukawa interactions; however, such terms would give singular soft theorems and are hence not allowed in this setting.} 
 On the other hand, the RHS of \reef{5ptsusyWI} will have a pole when $[24] \to 0$ (when momenta $p_2$ and $p_4$ go collinear), unless $A_5(Z \bar Z Z \bar Z Z)$  vanishes in that limit. One can explicitly check, using the expression \reef{5ptcomplexGal}, that it does not. Hence we conclude that 
\begin{quote}
{\em the quintic Galileon cannot be supersymmetrized while preserving the  Galileon symmetry \reef{galsym} for the complex scalar. }
\end{quote}
We will relax the condition of Galileon symmetry in Sec.~\ref{s:raxion}.

\subsection{Cubic Galileon}
The only possibly non-vanishing 3-scalar amplitude in any theory of massless scalars is constant, i.e.~it comes from $\phi^3$, and the resulting higher point amplitudes have singular soft limits. When a 3-particle amplitude vanishes, the associated cubic Lagrangian can be removed by a field redefinition. In particular, 
the cubic Galileon can be removed by a field redefinition of the form $\phi \to \phi + a (\pa \phi)^2$. 
This shuffles the information into $4$-, $5$-, and $6$-point interactions. There is no independent 6th order Galileon, so this means the cubic Galileon is equivalent to a particular choice of the quartic and quintic Galileon. This remains true also when there are multiple scalars. In particular, the quintic coupling will be non-zero. From the above, we immediately conclude that
\begin{quote}
{\em the cubic Galileon cannot be supersymmetrized while preserving the  Galileon symmetry \reef{galsym} for the complex scalar.}
\end{quote}
 
\subsection{Quartic Galileon \& supersymmetry, part 1}
\label{s:quartic1}
Soft subtracted recursion relations with $\sigma_Z=2$ show that there is a unique complex scalar quartic Galileon
whose amplitude is 
\be
 \label{A4gal4}
 A_4(Z \bar{Z} Z \bar{Z}) = g_4 \,s t u\,.
\ee
We have already introduced the supersymmetry Ward identity \reef{4ptsusyWI} that fixes all the other 4-point amplitudes in terms of this result. Thus the 4-particle sector is unique. Using the supersymmetry Ward identities, one can show \cite{ourlongpaper} 
that the soft behavior $\sigma_\psi$ of the fermion in the chiral multiplet is related to the scalar soft behavior as $\sigma_\psi = \sigma_Z$ OR $\sigma_\psi = \sigma_Z-1$. The recursion relations for 
$A_6(Z \bar{Z} Z \bar{Z} \psi \bar\psi)$ are valid in either case (by \reef{reccrit}). The condition of $a_i$-independence  passes for $\sigma_\psi =1$, but fails for $\sigma_\psi = \sigma_Z=2$.  
 The result that $\sigma_\psi =1$ then proves that a supersymmetrization of the quartic Galileon must have a regular shift symmetry for the fermions. 

Proceeding, the constructibility criterion \reef{reccrit} with $\sigma_\psi =1$ and $\sigma_Z=2$ shows that only amplitudes with at most a pair of fermions are constructible when based on  quartic interactions with coupling dimension $[g_4]=-6$. However, at 6-point order, we can exploit the supersymmetry Ward identities to fully construct all 6-particle amplitudes in the supersymmetric quartic Galileon theory. The supersymmetry Ward identities are
\be 
 \nonumber
  \begin{split}
  [25] A_6(Z \bar{Z}Z \bar{Z}Z \bar{Z})
  -[26] A_6(Z \bar{Z}Z \bar{Z}\psi \bar{\psi})
  +[24] A_6(Z \bar{Z}Z \bar{\psi}\psi \bar{Z}) 
  &=0\,\\
    [23] A_6(Z \bar{Z}Z \bar{\psi}\psi \bar{Z}) 
 +[25] A_6(Z \bar{Z}\psi \bar{\psi}Z \bar{Z})
  -[26] { A_6(Z \bar{Z}\psi \bar{\psi}\psi \bar{\psi})}
  &=0\,
\\
[31]  { A_6(Z \bar{\psi}\psi \bar{\psi}\psi \bar{Z}) }
 +[35]  { A_6(\psi \bar{\psi}\psi \bar{\psi}Z \bar{Z})}
  -[36] { A_6(\psi \bar{\psi}\psi \bar{\psi}\psi \bar{\psi})}
  &=0\,.
  \end{split}
\ee
We use the first identity to check that the amplitudes reconstructed with soft subtracted recursion relations are compatible with supersymmetry.
The second identity allows us to solve for the four-fermion amplitude, and with this result the third identity uniquely determines 
the 6-fermion amplitude. There are three more independent supersymmetry Ward identities: we use them as consistency checks to make sure all the 6-point amplitudes are compatible with the supersymmetry requirements. These checks all pass. We present the explicit results for the 6-particle amplitudes in \cite{ourlongpaper}.

At higher point, the constructible amplitudes with at most two fermions are not sufficient to solve the supersymmetry Ward identities. In a Lagrangian construction, there may therefore be an ambiguity starting at 8th orders in the fields in terms of independently supersymmetrizable operators which must not have any components with two fermions or less; such operators will involve so many derivatives that it is trivial that they can be compatible with the Galileon symmetry \reef{galsym} for the scalars  and shift symmetry for the fermions. 

Notice also that the constructible amplitudes satisfy the conservation of a $U(1)_R$ charge under which only the scalar $Z$ is charged. Such a symmetry is also 
respected by the supersymmetrization of DBI, but given the ambiguity in the non-constructible amplitudes the strongest statement we can say is that a supersymmetric quartic Galileon \textit{may} be consistent with such a symmetry.

In conclusion, 
\begin{quote}
 {\em we have found strong evidence for an $\mathcal{N}=1$ supersymmetrization of the  quartic Galileon. It is compatible with a Galileon symmetry \reef{galsym} for the complex scalar and shift symmetry for the fermion. It may not be a unique supersymmetrization, as there could be independently supersymmetrizable operators starting at 8th order in fields. }
\end{quote}
A superfield Lagrangian for $\mathcal{N}=1$ quartic Galileon was presented in \cite{Farakos:2013zya}. 
We find that (up to a sign) the 4-point amplitudes computed from \cite{Farakos:2013zya} agree with ours. Further comparisons are deferred to \cite{ourlongpaper}.

\subsection{Special Galileon vs.~supersymmetry}
The real scalar amplitudes resulting from \reef{A4gal4} are those of the special Galileon, which has the enhanced shift symmetry \reef{specgalsym}. However, this symmetry does not carry over to the complex scalar case (i.e.~it is broken by terms mixing the two scalars). This follows from using $\sigma=3$ in the soft subtracted recursion relations: this construction $A_6(Z \bar{Z}Z \bar{Z}Z \bar{Z})$ fails $a_i$-independence. We conclude that
\begin{quote}
 {\em for the quartic Galileon, the special Galileon symmetry \reef{specgalsym} is not compatible with supersymmetry.}
\end{quote}
This is what the argument based on the algebra indicated.

\section{Results II: Marrying Galileons and R-axions}
\label{s:raxion}
In this section, we find evidence for $\mathcal{N}=1$ supersym\-metric quartic and quintic Galileon theories in which the complex scalar $Z = (\phi + i \chi)/\!\sqrt 2$ bundles an honest Galileon $\phi$, who enjoys extended shift symmetry \reef{galsym}, with a second real scalar $\chi$, who only has regular shift symmetry. The second scalar $\chi$ is naturally identified as an R-axion and a scenario for this type of theory is partial supersymmetry breaking; this will be discussed further in \cite{ourlongpaper}.

\subsection{Quintic Galileon \& supersymmetry, part 2}
We start by writing the most general Ansatz for the amplitudes $A_5 ( Z \bar Z Z \bar Z Z)$, $A_5 ( Z \bar Z Z \bar \psi \psi )$, $A_5 ( Z \bar \psi \psi \bar \psi \psi )$ and their complex conjugates.  All other amplitudes must  be zero for a theory compatible with supersymmetry.  On this Ansatz of 122 free parameters we impose the following constraints:
\begin{itemize}
\item
Compatibility with supersymmetry via the supersymmetry Ward identities 
\be
\begin{aligned}
A_5 ( Z \bar Z Z \bar Z Z ) & = - \frac{[ 24 ]}{[ 25 ]} A_5 ( Z \bar Z Z \bar \psi \psi ) = \frac{[ 24 ]}{[ 35 ]} A_5 ( Z \bar \psi \psi \bar \psi \psi )\,, \\
A_5 ( \bar Z Z \bar Z Z \bar Z ) & = - \frac{\langle 24 \rangle}{\langle 25 \rangle} A_5 ( \bar Z Z \bar Z \psi \bar \psi ) = \frac{\langle 24 \rangle}{\langle 35 \rangle} A_5 ( \bar Z \psi \bar \psi \psi \bar \psi)\,.
\label{fivePtSUSY-WI}
\end{aligned}
\ee
\item
A shift symmetry for the complex field $Z$ in the form of $\sigma = 1$ soft behavior for the amplitudes of $Z$.
\item
Galileon symmetry for the real scalar field $\phi$ in the form of $\sigma = 2$ soft behavior imposed on the linear combinations of complex-scalar amplitudes, $$A_5 ( \phi \cdot \cdot \cdot \cdot ) = \frac{1}{\sqrt 2} \left ( A_5 ( Z \cdot \cdot \cdot \cdot ) + A_5 ( \bar Z \cdot \cdot \cdot \cdot ) \right )\,.$$ 
\end{itemize}
Imposing these constraints on our Ansatz left us with a  3-parameter 
family of solutions.  Interestingly, this solution comes with a `free' $\sigma = 1$ soft behavior for the fermions, that suggests that the theory is invariant under a shift of the fermions.  Moreover, the five-Galileon amplitude $A_5 ( \phi \phi \phi \phi \phi )$ matches the known real Galileon amplitude.

In order to further constrain the 
solution, we consider the 7-point amplitudes of  the 
DBI-Galileon theory.  The leading order contribution to these amplitudes is proportional to the product of the DBI coupling with mass dimension $-4$ and the the quintic Galileon coupling with mass dimension $-6$; it can be reconstructed using subtracted soft recursion relations with $\sigma_\phi = 2$, $\sigma_\chi = 1$ and $\sigma_\psi = 1$ if $2 n_\chi + n_f < 8$, where $n_\chi$ is the number of $\chi$-external states and $n_f$ is the the number of fermionic external states.  Demanding that the results of recursion are independent of the shift parameters $a_i$ \emph{uniquely} fixes the parameters of our solution.  The resulting scalar amplitude is
\be
\begin{aligned}
A_5 ( Z \bar Z Z \bar Z Z ) = s_{24} \Big [ 6 s_{24} s_{25} s_{45} + \Big ( 4 s_{12} s_{23} s_{45} + 2 s_{12} s_{24} s_{34} \\ + 2 s_{25}^2 s_{45} + s_{24} s_{25}^2 + ( 2 \leftrightarrow 4 ) \Big ) + ( 1 \leftrightarrow 5 ) + ( 3 \leftrightarrow 5 ) \Big ] - 4 s_{24}^4 \,,
\end{aligned}
\ee
while the amplitudes with fermions can be straightforwardly obtained from the supersymmetry Ward identities \reef{fivePtSUSY-WI}.

To conclude this section,
\begin{quote}
{\em we  find 
strong evidence for the existence of a supersymmetrization of the quintic Galileon.  In this theory, only one of the two scalar modes enjoys the full Galileon symmetry, while the second one, an R-symmetry axion has only a shift symmetry.}
\end{quote}

\subsection{Quartic Galileon \& supersymmetry, part 2}
A very similar analysis can be carried out for the quartic Galileon. The 4-point scalar amplitude has two independent terms, $A_4 ( Z \bar Z Z \bar Z ) = r_1 \,stu + r_2\, t^3$. When $r_2 = 0$, we recover the quartic Galileon from Sec.~\ref{s:quartic1} which has $\sigma=2$ for the complex scalar. Computing all constructible (by \reef{reccrit}) 6-point amplitudes in both the decoupled 
Galileon and DBI-Galileon places no constraints on the couplings $r_1$ and $r_2$. {\em This is evidence that there may exist a 2-parameter family of quartic $\mathcal{N}=1$ supersymmetric Galileons in which the complex scalar is composed of a galileon and an R-axion}. Note that if one were to try the same for DBI, one would find that the R-axion automatically has its symmetry enhanced to \reef{galsym}; we comment further on that interesting result in \cite{ourlongpaper}.

\section*{Acknowledgments}

We would like to thank Ratin Akhoury, Clifford Cheung, Nima Arkani-Hamed, Yu-tin Huang, and Chia-Hsien Shen for useful discussions.
All four authors are grateful to the Kavli Institute for Theoretical Physics, UC Santa Barbara, for  
hospitality during the `Scattering Amplitudes and Beyond' program.
This work was supported in part by the US
Department of Energy under Grant No.~de-sc0007859.






\section*{References}
\bibliographystyle{unsrt}
\bibliography{susyGalBib}

\begin{thebibliography}{10}

\bibitem{Dvali:2000hr}
G.~R. Dvali, Gregory Gabadadze, and Massimo Porrati.
\newblock {4-D gravity on a brane in 5-D Minkowski space}.
\newblock {\em Phys. Lett.}, B485:208--214, 2000.

\bibitem{Nicolis:2008in}
Alberto Nicolis, Riccardo Rattazzi, and Enrico Trincherini.
\newblock {The Galileon as a local modification of gravity}.
\newblock {\em Phys. Rev.}, D79:064036, 2009.

\bibitem{deRham:2010ik}
Claudia de~Rham and Gregory Gabadadze.
\newblock {Generalization of the Fierz-Pauli Action}.
\newblock {\em Phys. Rev.}, D82:044020, 2010.

\bibitem{deRham:2010kj}
Claudia de~Rham, Gregory Gabadadze, and Andrew~J. Tolley.
\newblock {Resummation of Massive Gravity}.
\newblock {\em Phys. Rev. Lett.}, 106:231101, 2011.

\bibitem{deRham:2010eu}
Claudia de~Rham and Andrew~J. Tolley.
\newblock {DBI and the Galileon reunited}.
\newblock {\em JCAP}, 1005:015, 2010.

\bibitem{Goon:2010xh}
Garrett~L. Goon, Kurt Hinterbichler, and Mark Trodden.
\newblock {Stability and superluminality of spherical DBI galileon solutions}.
\newblock {\em Phys. Rev.}, D83:085015, 2011.

\bibitem{Goon:2011qf}
Garrett Goon, Kurt Hinterbichler, and Mark Trodden.
\newblock {Symmetries for Galileons and DBI scalars on curved space}.
\newblock {\em JCAP}, 1107:017, 2011.

\bibitem{Adams:2006sv}
Allan Adams, Nima Arkani-Hamed, Sergei Dubovsky, Alberto Nicolis, and Riccardo
  Rattazzi.
\newblock {Causality, analyticity and an IR obstruction to UV completion}.
\newblock {\em JHEP}, 10:014, 2006.

\bibitem{Cheung:2014dqa}
Clifford Cheung, Karol Kampf, Jiri Novotny, and Jaroslav Trnka.
\newblock {Effective Field Theories from Soft Limits of Scattering Amplitudes}.
\newblock {\em Phys. Rev. Lett.}, 114(22):221602, 2015.

\bibitem{Hinterbichler:2015pqa}
Kurt Hinterbichler and Austin Joyce.
\newblock {Hidden symmetry of the Galileon}.
\newblock {\em Phys. Rev.}, D92(2):023503, 2015.

\bibitem{Farakos:2013zya}
Fotis Farakos, Cristiano Germani, and Alex Kehagias.
\newblock {On ghost-free supersymmetric galileons}.
\newblock {\em JHEP}, 11:045, 2013.

\bibitem{Koehn:2013hk}
Michael Koehn, Jean-Luc Lehners, and Burt Ovrut.
\newblock {Supersymmetric cubic Galileons have ghosts}.
\newblock {\em Phys. Rev.}, D88(2):023528, 2013.

\bibitem{Goon:2012dy}
Garrett Goon, Kurt Hinterbichler, Austin Joyce, and Mark Trodden.
\newblock {Galileons as Wess-Zumino Terms}.
\newblock {\em JHEP}, 06:004, 2012.

\bibitem{Roest:2017uga}
Diederik Roest, Pelle Werkman, and Yusuke Yamada.
\newblock {Internal Supersymmetry and Small-field Goldstini}.
\newblock 2017.

\bibitem{Bagger:1994vj}
J.~Bagger and A.~Galperin.
\newblock {Matter couplings in partially broken extended supersymmetry}.
\newblock {\em Phys. Lett.}, B336:25--31, 1994.

\bibitem{Toppan:2001qb}
Francesco Toppan.
\newblock {On anomalies in classical dynamical systems}.
\newblock {\em J. Nonlin. Math. Phys.}, 8:518--533, 2001.

\bibitem{PhysRevLett.16.408.2}
E.~Fabri and L.~E. Picasso.
\newblock Quantum field theory and approximate symmetries.
\newblock {\em Phys. Rev. Lett.}, 16:408--410, Mar 1966.

\bibitem{Dumitrescu:2011iu}
Thomas~T. Dumitrescu and Nathan Seiberg.
\newblock {Supercurrents and Brane Currents in Diverse Dimensions}.
\newblock {\em JHEP}, 07:095, 2011.

\bibitem{Coleman:1969sm}
Sidney~R. Coleman, J.~Wess, and Bruno Zumino.
\newblock {Structure of phenomenological Lagrangians. 1.}
\newblock {\em Phys. Rev.}, 177:2239--2247, 1969.

\bibitem{Callan:1969sn}
Curtis~G. Callan, Jr., Sidney~R. Coleman, J.~Wess, and Bruno Zumino.
\newblock {Structure of phenomenological Lagrangians. 2.}
\newblock {\em Phys. Rev.}, 177:2247--2250, 1969.

\bibitem{Volkov:1973vd}
Dmitri~V. Volkov.
\newblock {Phenomenological Lagrangians}.
\newblock {\em Fiz. Elem. Chast. Atom. Yadra}, 4:3--41, 1973.

\bibitem{ourlongpaper}
Henriette Elvang, Marios Hadjiantonis, Callum R.~T. Jones, and Shruti
  Paranjape.
\newblock {to appear}.

\bibitem{Elvang:2013cua}
Henriette Elvang and Yu-tin Huang.
\newblock {Scattering Amplitudes}.
\newblock 2013.

\bibitem{Elvang:2015rqa}
Henriette Elvang and Yu-tin Huang.
\newblock {\em {Scattering Amplitudes in Gauge Theory and Gravity}}.
\newblock Cambridge University Press, 2015.

\bibitem{Elvang:2016qvq}
Henriette Elvang, Callum R.~T. Jones, and Stephen~G. Naculich.
\newblock {Soft Photon and Graviton Theorems in Effective Field Theory}.
\newblock {\em Phys. Rev. Lett.}, 118(23):231601, 2017.

\bibitem{Kampf:2013vha}
Karol Kampf, Jiri Novotny, and Jaroslav Trnka.
\newblock {Tree-level Amplitudes in the Nonlinear Sigma Model}.
\newblock {\em JHEP}, 05:032, 2013.

\bibitem{Cheung:2015cba}
Clifford Cheung, Chia-Hsien Shen, and Jaroslav Trnka.
\newblock {Simple Recursion Relations for General Field Theories}.
\newblock {\em JHEP}, 06:118, 2015.

\bibitem{Cheung:2015ota}
Clifford Cheung, Karol Kampf, Jiri Novotny, Chia-Hsien Shen, and Jaroslav
  Trnka.
\newblock {On-Shell Recursion Relations for Effective Field Theories}.
\newblock {\em Phys. Rev. Lett.}, 116(4):041601, 2016.

\bibitem{Cheung:2016drk}
Clifford Cheung, Karol Kampf, Jiri Novotny, Chia-Hsien Shen, and Jaroslav
  Trnka.
\newblock {A Periodic Table of Effective Field Theories}.
\newblock {\em JHEP}, 02:020, 2017.

\bibitem{Luo:2015tat}
Hui Luo and Congkao Wen.
\newblock {Recursion relations from soft theorems}.
\newblock {\em JHEP}, 03:088, 2016.

\bibitem{Hinterbichler:2010xn}
Kurt Hinterbichler, Mark Trodden, and Daniel Wesley.
\newblock {Multi-field galileons and higher co-dimension branes}.
\newblock {\em Phys. Rev.}, D82:124018, 2010.

\end{thebibliography}

%
%

\end{document}